# On the Design of Efficient CSMA Algorithms for Wireless Networks

J. Ghaderi and R. Srikant

Department of ECE and Coordinated Science Lab.

University of Illinois at Urbana-Champaign

{jghaderi, rsrikant}@illinois.edu

**Abstract**

Recently, it has been shown that CSMA algorithms which use queue length-based link weights can achieve throughput optimality in wireless networks. In particular, a key result by Rajagopalan, Shah, and Shin (2009) shows that, if the link weights are chosen to be of the form $\log\log(q)$ (where $q$ is the queue-length), then throughput optimality is achieved. In this paper, we tighten their result by showing that throughput optimality is preserved even with weight functions of the form $\log(q)/g(q)$, where $g(q)$ can be a function that increases arbitrarily slowly. The significance of the result is due to the fact that weight functions of the form $\log(q)/g(q)$ seem to achieve the best delay performance in practice.

## I. INTRODUCTION

Efficient operation of wireless networks has always been a difficult task due to the inherent broadcast nature of the wireless medium. Transmission by a user can cause an interference for its neighbors. If two neighboring users transmit at the same time, the *Signal-to-Noise-plus-Interference Ratio* (SINR) of the users' links could go below the required SINR for the successful decoding of data packets at their corresponding receivers. In this case, we say that their messages *collide* with each other. Therefore, multiple users can transmit at the same time provided that they do not cause significant interference for each other. The users need a distributed *Medium Access Control* (MAC) protocol to determine which users should transmit which makes the optimal operation even harder.

CSMA (Carrier sense Multiple access) type protocols are an important class of MAC protocols due to their simplicity of implementation, and have been widely used in practice. e.g., in WLANs (IEEE 802.11 Wi-Fi) or emerging wireless mesh networks. In these protocols, each user listens to the channel and can transmit, with some probability, only when the channel is not busy. Despite the extreme simplicity of the CSMA-type algorithms, their efficiency have been always questionable. In this paper, we consider efficient design of such CSMA-type algorithms that can achieve maximum throughput and good delay performance.

The wireless network can be modeled by its conflict graph (or interference model), where two communication links form two neighboring nodes in the conflict graph, if they cannot transmit simultaneously. The well-known result of Tassiulas and Ephremides [1] states that the Maximum Weight Scheduling (MWS) algorithm is throughput optimal, where weights are queue-lengths. However, for a general network, MWS involves finding the maximum weight independent set of the conflict graph in each time slot which is a formidable task, and hence, is not implementable. This has led to a rich amount of literature on design of approximate algorithms to alleviate the computational complexity of the MWS algorithm. A recent result in this direction of research can be found in [9].

Recently, it has been shown that it is possible to design CSMA algorithms that are throughput-optimal. Reference [10] develops an algorithm that adaptively chooses the CSMA parameters under a time-scale separation assumption, i.e., the CSMA Markov chain converges to its stationary distribution instantaneously compared to the time-scale of adaptation of the CSMA parameters. This time-scale separation assumption was later verified by a stochastic approximation type argument [11], [12]. In particular, an important recent work by Rajagopalan, Shah, and Shin [4] builds an algorithm upon a Metropolis-Hastings sampling mechanism (Glauber dynamics over the set of feasible schedules) along with selection of link weights to be of the form $\log\log(q)$ ($q$ is the queue-length). To establish the efficiency of the algorithm, they present a novel adiabatic-like theorem for the underlying queueing network: by choosing the weights to be of the form $\log\log q$, the underlying Markov chain behaves in an adiabatic manner such that it remains close to its equilibrium distribution. Similar algorithms with fixed link weights were developed earlier in [8] and [13].

Although a weight function of the form $\log\log q$ stabilizes the network, the resulting scheduling algorithm reacts very slowly to changes in queue lengths which, in turn, results in a poor delay performance. In this paper, we show that, by choosing weights to be of the form $\log q/g(q)$, a network adiabatic property still holds. As a result, we prove that the CSMA algorithms with such weight functions are throughput optimal. The function $g$ can grow arbitrarily slowly such that $\log q/g(q)$ behaves very similarly to $\log(q)$. The significance of the result is due to the fact that such weight functions seem to also achieve the best delay performance in practice.

The remainder of the paper is organized as follows. In section II, we briefly describe our model of wireless networks. The main results of the paper are presented in section III. Section IV is devoted to the proofs. Some simulation results are presented in section V. Finally, we will end the paper with some concluding remarks.

## II. Model of Wireless Network

Consider a set of nodes where each node could be a source and/or a destination for another source. For now, we assume a single hop communication scheme but the results are naturally extendable to the multihop case. Therefore, there are $N$ communication links, each of which corresponds to a source-destination pair.

Time is slotted and arrival process to each link is assumed to be discrete-time, where $a_l(t)$ is the number of packets arriving at link $l$ in time slot $t$. For simplicity, assume that $\{a_l(t)\}_{t=0}^{\infty}$, for $l = 1, .., N$, are independent Bernoulli processes with parameter $\lambda = [\lambda_l; l = 1, .., N]$. In each time slot, one packet could be successfully transmitted over a link if the link Signal-to-Interference-plus-Noise-Ratio (SINR) is high enough. We use the notion of the *conflict graph* to capture the interference constraints or technological ones[1]. Let $G(V, E)$ denote the conflict graph of the wireless network, where each node in the conflict graph is a communication link in the wireless network. There is an edge $(l, k) \in E$ between nodes $l$ and $k$ if simultaneous transmissions over communication links $l$ and $k$ are not successful. Therefore, at each time slot, the active links should form an independent set of $G$, i.e., no two scheduled nodes can share an edge in $G$. Formally, a schedule can be represented by a vector $X = [x_s : s = 1, ..., N]$ such that $x_s \in \{0, 1\}$ and $x_i + x_j \leq 1$ for all $(i, j) \in E$. let $\mathcal{M}$ denote the set of all feasible schedules.

Each link $l$ is associated with a queue $q_l$, where the queue dynamics are given by

$$q_l(t) = (q(t-1) - x_l(t))^+ + a_l(t)$$

[1]For example, a node cannot transmit and receive at the same time.

for $t \geq 0$ and $l = 1, ..., N$. The vector of queue lengths is denoted by $q(t) = [q_l(t) : l = 1, ..., N]$.

A scheduling algorithm is a policy to determine which schedule to be used in each time slot. The capacity region of the network is defined to be the set of all arrival rates $\lambda$ that can be supported by the network, i.e., for which there exists a scheduling algorithm that can stabilize the queues. It is known, e.g.[1], that the capacity region is given by

$$\Lambda = \{\lambda \geq 0 : \exists \mu \in \text{Co}(\mathcal{M}),\ \lambda < \mu\}$$

where $\text{Co}(.)$ is the convex hull operator. When dealing with vectors, inequalities are interpreted component-wise.

A scheduling algorithm is throughput-optimal if it can stabilize the network for any arrival rate in $\Lambda$. An important class of the throughput-optimal algorithms is the *maximum-weight scheduling* (MWS) algorithm where at each time slot $t$, the scheduling decision $\rho(t)$ satisfies

$$\rho(t) = \arg \max_{X \in \mathcal{M}} \sum_{l=1}^{N} x_l w_l(t).$$

where $w_l(t)$ is the weight of link $l$ at time slot $t$. In [1], it was proved that the MWS algorithm is throughput-optimal for $w_l(t) = q_l(t)$. A natural generalization of the MWS algorithm in [2] uses a weight $f(q_l(.))$ instead of $q_l(.)$ with the following properties.

1) $f : [0, \infty] \to [0, \infty]$ is a nondecreasing continuous function with $\lim_{q_l \to \infty} f(q_l) = \infty$.
2) Given any $M_1, M_2 > 0$, and $0 < \epsilon < 1$, there must exist a $Q < \infty$ such that for $q_l > Q$:

$$(1 - \epsilon) f(q_l) \leq f(q_l - M_1) \leq f(q_l + M_2) \leq (1 + \epsilon) f(q_l)$$

**Lemma 1.** *Suppose $f$ is a strictly concave and monotonically increasing function, with $f(0) = 0$, then it satisfies the conditions (1) and (2) above.*

See the appendix for the proof. In this paper, we use a function $f$ with properties of Lemma 1.

## III. Main Result

### A. Basic Algorithm

For our algorithm, we choose the wight of link $l$ to be

$$\widetilde{w}_l(t) = \max\left(w_l(t), w_{min}(t)\right) \tag{1}$$

where

$$w_l(t) = f(q_l(t)), \tag{2}$$

$$w_{min}(t) = \frac{\epsilon}{2N} f(q_{max}(t)), \tag{3}$$

$$f(x) = \frac{\log(1+x)}{g(x)}, \tag{4}$$

and $q_{max}(t)$ is the length of the largest queue in the network at time $t$ and assumed to be known. The function $g(x)$ is a strictly increasing function chosen such that $f$ is a strictly concave increasing function, for example $g(x) = \log(e + \log(1+x))$ or $g(x) = (\log(1+x))^{\theta}$ for some $0 < \theta < 1$. In this paper, all $\log$'s are in base $e$

Consider the conflict graph $G(V, E)$ of the network as defined earlier. Denote the neighbors of $i$ by a set $\mathcal{N}(i) = \{k \in V : (i,k) \in E\}$. At each time slot $t$, a node $i$ is chosen uniformly at random, with probability $\frac{1}{N}$, then

(i) If $x_j(t-1) = 0$ for all nodes $j \in \mathcal{N}(i)$, then $x_i(t) = 1$ with probability $\frac{\exp(\widetilde{w}_i(t))}{1+\exp(\widetilde{w}_i(t))}$, and $x_i(t) = 0$ with probability $\frac{1}{1+\exp(\widetilde{w}_i(t))}$ .
Otherwise, $x_i(t)$=0.

(iii) $x_j(t) = x_j(t-1)$ for all $j \neq i$.

The following theorem states the main result regarding the throughput optimality of the algorithm.

**Theorem 1.** *Consider any $\epsilon > 0$. The algorithm can stabilize the network for any $\lambda \in (1-\epsilon)\Lambda$, if the weight function is chosen to be in the form of $f(x) = \frac{\log(1+x)}{g(x)}$. The function $g(x)$ is a strictly increasing function chosen such that $f$ is a strictly concave increasing function. In particular, the algorithm with the following weight functions is throughput-optimal: $f(x) = \frac{\log(1+x)}{\log(e+\log(1+x))}$ or $f(x) = (\log(1+x))^{1-\theta}$ for some $0 < \theta < 1$.*

### *B. Distributed Implementation*

The basic algorithm is based on Glauber-Dynamics with one site update at each time. For distributed implementation, we need a randomized mechanism to select a link uniformly at each time slot. We use the Q-CSMA idea [3] to perform the link selection as follows. Each time slot is divided into a control slot and a data slot. The control slot, which is much smaller than the data slot, is used to generate a transmission schedule for the data slot. First, the network selects a set of links $m(t)$ that do not conflict with each other. Then, it performs the Glauber-Dynamics updates, in parallel, over links $m(t)$ to produce a transmission schedule $X(t)$ for data transmission. $m(t)$ is called the decision schedule at time $t$. For example, a simple randomized mechanism to generate $m(t)$ is as follows. In control slot $t$, each link $l$ sends an INTENT message with probability $1/2$. If $l$ does not hear any INTENT messages from its neighboring links $\mathcal{N}(l)$, it will be included in $m(t)$, otherwise it will not be included in $m(t)$. Therefore, by the end of the control slot, any feasible decision schedule $m(t) \subseteq \mathcal{M}$ could be selected with a positive probability $\alpha(m(t))$. Once a link knows whether it is included in the decision schedule, it can determine its state in the data slot based on its carrier sensing information (i.e., whether its conflicting links were active in the previous data slot) and the activation probability for the current slot (based on its queue length).

To determine the weight at each link $l$, $q_{max}(t)$ is needed. Instead, each link $l$ can maintain an estimate of $q_{max}(t)$. We can use the procedure suggested in [4] to estimate $q_{max}(t)$, and use Lemma 2 of [4] to complete the stability proof. So we do not pursue this issue here. In practical networks $\frac{\epsilon}{2N}\log(1+q_{max})$ is small and we can use the weight function $f$ directly, and thus, there may not be any need to know $q_{max}(t)$.

**Corollary 1.** *Under the weight function $f$ specified in Theorem 1, the distributed algorithm can stabilize the network for any $\lambda \in (1-\epsilon)\Lambda$.*

## IV. Proof of the Main result

Before we start the proof, some preliminaries, regarding stationary distribution and mixing time of Glauber dynamics, are needed.

### A. Preliminaries

Consider a time-homogenous discrete-time Markov chain over the finite state-space $\mathcal{M}$. For simplicity, we index the elements of $\mathcal{M}$ by $1, 2, ..., r$, where $r = |\mathcal{M}|$. Assume the Markov chain is irreducible and aperiodic, so that a unique stationary distribution $\pi = [\pi(1), ..., \pi(r)]$ always exists.

*1) Distance between probability distributions:* First, we introduce two convenient norms on $\mathbb{R}^r$ that are linked to the stationary distribution. Let $\ell^2(\pi)$ be the real vector space $\mathbb{R}^r$ endowed

with the scalar product

$$\langle z, y\rangle_\pi = \sum_{i=1}^{r} z(i)y(i)\pi(i).$$

Then, the norm of $z$ with respect to $\pi$ is defined as

$$\|z\|_\pi = \left(\sum_{i=1}^{r} z(i)^2\pi(i)\right)^{1/2}.$$

We shall also use $\ell^2(\frac{1}{\pi})$, the real vector space $\mathbb{R}^r$ endowed with the scalar product

$$\langle z, y\rangle_{\frac{1}{\pi}} = \sum_{i=1}^{r} z(i)y(i)\frac{1}{\pi(i)}$$

and its corresponding norm. For any two strictly positive probability vectors $\mu$ and $\pi$, the following relationship holds

$$\|\mu - \pi\|_{\frac{1}{\pi}} = \|\frac{\mu}{\pi} - 1\|_\pi \geq 2\|\mu - \pi\|_{TV},$$

where $\|\pi - \mu\|_{TV}$ is the total variation distance

$$\|\pi - \mu\|_{TV} = \frac{1}{2}\sum_{i=1}^{r} |\pi(i) - \mu(i)|.$$

*2) Glauber dynamics:* Consider a graph $G(V, E)$. Glauber dynamics is a Markov chain to generate the independent sets of $G$. So, the state space $\mathcal{M}$ consists of all independent sets of $G$. Let $|V| = N$. Given a weight vector $\tilde{W} = [\tilde{w}_1, \tilde{w}_2, ..., \tilde{w}_N]$, at each time $t$, a node $i$ is chosen uniformly at random, with probability $\frac{1}{N}$, then

(i) If $x_j(t-1) = 0$ for all nodes $j \in \mathcal{N}(i)$, then $x_i(t) = 1$ with probability $\frac{\exp(\tilde{w}_i)}{1+\exp(\tilde{w}_i)}$, or $x_i(t) = 0$ with probability $\frac{1}{1+\exp(\tilde{w}_i)}$ .

Otherwise, $x_i(t)$=0.

(iii) $x_j(t) = x_j(t-1)$ for all $j \neq i$.

The corresponding Markov chain is irreducible, aperiodic, and reversible over $\mathcal{M}$, and its stationary distribution is given by

$$\pi(\rho) = \frac{1}{Z}\exp(\sum_{i\in\rho} \tilde{w}_i); \quad \rho \in \mathcal{M}, \tag{5}$$

where $Z$ is the normalizing constant.

The basic algorithm uses a time-varying version of the above Glauber dynamics, where the weights change with time. This yields a time-inhomogeneous Markov chain but we will see that, for the proper choice of weights, it behaves similarly to the Glauber dynamics.

*3) Mixing time of Glauber dynamics:* The convergence to steady state distribution is geometric with a rate equal to the *second largest eigenvalue modulus* (SLEM) of the transition matrix as it is described next [6].

**Lemma 2.** *Let $P$ be an irreducible, aperiodic, and reversible transition matrix on the finite state space $\mathcal{M}$ with the stationary distribution $\pi$. Then, the eigenvalues of $P$ are ordered in such a way that*

$$\lambda_1 = 1 > \lambda_2 \geq ... \geq \lambda_r > -1,$$

*and for any initial probability distribution $\mu_0$ on $\mathcal{M}$, and for all $n \geq 1$*

$$\|\mu_0 \mathbf{P}^n - \pi\|_{\frac{1}{\pi}} \leq \sigma^n \|\mu_0 - \pi\|_{\frac{1}{\pi}}, \tag{6}$$

*where $\sigma = \max\{\lambda_2, |\lambda_r|\}$ is the SLEM of $P$.*

The following Lemma gives an upper bound on the SLEM $\sigma(P)$ of Glauber dynamics.

**Lemma 3.** *For the Glauber Dynamics with the weight vector $\tilde{W}$ on a graph $G(V, E)$ with $|V| = N$,*

$$\sigma \leq 1 - \frac{1}{16^N \exp(4N\tilde{w}_{max})},$$

*where $\tilde{w}_{max} = \max_{i \in V} \tilde{w}_i$.*

We define *the mixing time* as $T = \frac{1}{1-\sigma}$, so

$$T \leq 16^N \exp(4N\tilde{w}_{max}) \tag{7}$$

Simple calculation, based on Lemma 2, reveals that the amount of time needed to get close to the stationary distribution is proportional to $T$.

### *B. A key lemma*

At any time slot $t$, given the weight vector $\tilde{W}(t) = [\tilde{w}_1(t), ..., \tilde{w}_N(t)]$, the MWS algorithm should solve

$$\max_{\rho \in \mathcal{M}} \sum_{i \in \rho} \tilde{w}_i(t),$$

instead, our algorithm tries to simulate a distribution

$$\pi_t(\rho) = \frac{1}{Z} \exp(\sum_{i \in \rho} \tilde{w}_i(t)); \quad \rho \in \mathcal{M}, \tag{8}$$

i.e., the stationary distribution of Glauber dynamics with the weight vector $\tilde{W}(t)$ at time $t$.

Let $P_t$ denote the transition probability matrix of Glauber dynamics with the weight vector $\tilde{W}(t)$. Also let $\mu_t$ be the true probability distribution of the inhomogeneous-time chain, over the set of schedules $\mathcal{M}$, at time $t$. Therefore, we have $\mu_t = \mu_{t-1}P_t$. Let $\pi_t$ denote the stationary distribution of the time-homogenous Markov chain with $P = P_t$ as in (8). By choosing proper $w_{min}$ and $f(.)$, we aim to ensure that $\mu_t$ and $\pi_t$ are close enough, i.e.,

$$\|\pi_t - \mu_t\|_{TV} \leq \delta/4$$

for some $\delta$ arbitrary small.

Let $w_{max}(t) = f(q_{max}(t))$. The following lemma gives a sufficient condition under which the probability distribution of the inhomogeneous Markov chain is close to the stationary distribution of the homogenous chain.

**Lemma 4.** *Given any $\delta > 0$, $\|\pi_t - \mu_t\|_{TV} \leq \frac{\delta}{4}$ holds for all $t \geq t^*$, if*

$$\alpha_t T_{t+1} \leq \delta/16 \text{ for all } t > 0, \tag{9}$$

*where*

(i) $\alpha_t = 2Nf'(f^{-1}(w_{min}(t+1)) - 1)$,

(ii) *$t^*$ is the smallest $t$ such that*

$$\sum_{k=1}^{t} \frac{1}{T_k^2} \geq \ln(4/\delta) + N(w_{max}(0) + \log 2)/2, \tag{10}$$

*and $T_{t+1}$ is the mixing time of the Glauber dynamics with the weight vector $\tilde{W}(t+1)$.*

Lemma 4 states a condition under which $\|\pi_t - \mu_t\|_{TV} \leq \frac{\delta}{4}$ for all $t \geq t^*$. Instead, assume that (9) holds only when $\|q(t)\| \geq q_{th}$[2] for a constant $q_{th} > 0$. Let $t_1$ be the first time that $\|q(t)\|$ hits $q_{th}$. Then, after that, it takes $t^*$ time slots for the chain to get close to $\pi_t$ if $\|q(t)\|$ remains above $q_{th}$ for $t_1 \leq t \leq t_1 + t^*$. Alternatively, we can say that $\|\pi_t - \mu_t\|_{TV} \leq \frac{\delta}{4}$ if $\|q(t)\| \geq q_{th} + t^*$ since at each time slot at most one departure can happen and this guarantees that $\|q(t)\| \geq q_{th}$ for, at least, the past $t^*$ time slots. This immediately implies the following Lemma that we will use in the proof of the main result.

[2]In this paper, $\|q(t)\| = \|q(t)\|_\infty = \max_i q_i(t) = q_{max}(t)$.

**Lemma 5.** *Given any* $\delta > 0$, $\|\pi_t - \mu_t\|_{TV} \leq \frac{\delta}{4}$ *holds when* $\|q(t)\| \geq q_{th} + t^*$, *if there exists a* $q_{th}$ *such that*

$$\alpha_t T_{t+1} \leq \delta/16 \text{ whenever } \|q(t)\| > q_{th}, \tag{11}$$

*where*

(i) $\alpha_t = 2Nf'(f^{-1}(w_{min}(t+1)) - 1)$

(ii) $T_t \leq 16^N \exp(4Nw_{max}(t))$

(ii) $t^*$ *is the smallest* $t$ *such that*

$$\sum_{k=t_1:\|q(t_1)\|=q_{th}}^{t_1+t^*} \frac{1}{T_k^2} \geq \ln(4/\delta) + N(f(q_{th}) + \log 2)/2. \tag{12}$$

In the above Lemma, condition $(ii)$ is based on the upper bound of (7) and the fact that $\tilde{w}_{max}(t) = w_{max}(t)$.

In other words, Lemma 5 states that when queue lengths are large, the observed distribution of the schedules is close to the desired stationary distribution.

**Remark 1.** *We will later see that, to satisfy condition (11) and to find a finite* $t^*$ *satisfying (12) in Lemma 5, the function* $f(.)$ *cannot be faster than* $\log(.)$. *In fact, the function* $f$ *must be slightly slower than* $\log(.)$ *to make the weight dynamics slow enough such that the distribution of the schedules remains close to the stationary distribution.*

**Remark 2.** *The above Lemma is a generalization of Lemma* 12 *(Network Adiabatic Theorem) of [4]. Here we consider general functions* $f(.)$, *whereas [4] considers a particular function* $\log\log(.)$. *The generalization allows us to use functions which are close to* $\log(.)$ *and perform much better than* $\log\log(.)$ *in simulations. The proof of Lemma 4 is presented in the appendix.*

### *C. Throughput optimality*

We will use the following Lemma [2] to prove the throughput-optimality of the algorithm.

**Lemma 6.** *For a scheduling algorithm, if given any* $0 < \epsilon < 1$ *and* $0 < \delta$*<1, there exists a* $B(\delta, \epsilon) > 0$ *such that: in any time slot* $t$, *with probability larger than* $1 - \delta$, *the scheduling algorithm chooses a schedule* $X(t) \in \mathcal{M}$ *that satisfies*

$$\sum_{i \in X(t)} w_i(t) \geq (1-\epsilon) \max_{\rho \in \mathcal{M}} \sum_{i \in \rho} w_i(t)$$

*whenever* $\|\mathbf{q}(t)\| > B(\delta, \epsilon)$*, then the scheduling algorithm is throughput-optimal.*

**Remark 3.** *Throughput optimality in Lemma 6 means that, for all the rates inside the capacity region, system will be stable in the mean (See [2] for more details), i.e.,*

$$\limsup_{T\to\infty} \frac{1}{T}\sum_{t=0}^{T-1} E\left[\left(\sum_{i=1}^{N} f^2(q_i(t))\right)^{\frac{1}{2}}\right] < \infty. \tag{13}$$

*In our setting, the queuing system is an irreducible and aperiodic Markov chain, and therefore stability-in-the mean property (13) implies that the Markov chain is also positive recurrent [7] .*

Let $w^*(t) = \max_{\rho\in\mathcal{M}} \sum_{i\in\rho} w_i(t)$. Let us define the following set:

$$\chi_t = \{\rho \in \mathcal{M} : \sum_{i\in\rho} w_i(t) < (1-\epsilon)w^*(t)\}$$

Therefore, we need to show that

$$\mu_t(\chi_t) = \sum_{\rho\in\chi_t} \mu_t(\rho) \leq \delta$$

for $\|q(t)\|$ large enough. Suppose $f(.)$ and $w_{min}$ are chosen such that $\alpha_t T_{t+1} \leq \delta/16$ whenever $\|q(t)\| > q_{th}$ for some constant $q_{th} > 0$ to be determined later. Then, it follows from Lemma 5 that whenever $\|q(t)\| > q_{th} + t^*$,

$$2\|\mu_t - \pi_t\|_{TV} \leq \delta/2,$$

and consequently,

$$\sum_{\rho\in\mathcal{M}} |\mu_t(\rho) - \pi_t(\rho)| \leq \delta/2.$$

Thus,

$$\begin{aligned} |\sum_{\rho\in\chi_t}(\mu_t(\rho) - \pi_t(\rho))| &\leq \sum_{\rho\in\chi_t} |\mu_t(\rho) - \pi_t(\rho)| \\ &\leq \delta/2 \end{aligned}$$

which yields

$$\sum_{\rho\in\chi_t} \mu_t(\rho) \leq \sum_{\rho\in\chi_t} \pi_t(\rho) + \delta/2.$$

Therefore, to ensure that $\sum_{\rho\in\chi_t} \mu_t(\rho) \leq \delta$, it suffices to have

$$\sum_{\rho\in\chi_t} \pi_t(\rho) \leq \delta/2.$$

But

$$\sum_{\rho \in \chi_t} \pi_t(\rho) \quad = \quad \sum_{\rho \in \chi_t} \frac{1}{Z_t} \exp(\sum_{i \in \rho} \widetilde{w}_i(t))$$

where

$$\widetilde{w}_i(t) = \max\{w_i(t) \ , \ w_{min}(t)\} \le w_i(t) + w_{min}(t).$$

So

$$\begin{aligned}
\sum_{\rho \in \chi_t} \pi_t(\rho) &\le \sum_{\rho \in \chi_t} \frac{1}{Z_t} \exp(\sum_{i \in \rho}(w_i(t) + w_{min}(t))) \\
&= \sum_{\rho \in \chi_t} \frac{1}{Z_t} \exp(\sum_{i \in \rho} w_i(t)) \exp(|\rho| w_{min}(t)) \\
&\le \sum_{\rho \in \chi_t} \frac{1}{Z_t} \exp((1-\epsilon) w^*(t)) \exp(N w_{min}(t))
\end{aligned}$$

and

$$Z_t = \sum_{\rho \in \mathcal{M}} \exp(\sum_{i \in \rho} \widetilde{w}_i(t)) > \sum_{\rho \in \mathcal{M}} \exp(\sum_{i \in \rho} w_i(t)) > e^{w^*(t)}.$$

Therefore,

$$\sum_{\rho \in \chi_t} \pi_t(\rho) \quad \le \quad 2^N \exp(N w_{min}(t) - \epsilon w^*(t))$$

and $w^*(t) \ge w_{max}(t)$. So, it suffices to have

$$2^N \exp(N w_{min}(t) - \epsilon w_{max}(t)) \le \delta/2$$

when $\|q(t)\| > q_{th} + t^*$. The choice of $w_{min}(t) = \frac{\epsilon}{2N} w_{max}(t)$, satisfies the above condition for $\|q(t)\| > B$, where

$$B = \max\left\{ q_{th} + t^*, f^{-1}\left( \frac{N \log 2 + \log \frac{2}{\delta}}{\epsilon/2} \right) \right\}. \tag{14}$$

*D. A class of weight functions with the maximum throughput property*

In this section, we describe a family of weight functions $f$ that yield a maximum throughput algorithm.

The function $f$ needs to satisfy Lemma 5. Roughly speaking, since the mixing time $T$ is exponential in $w_{max}$, $f'(f^{-1}(w_{min}))$ must be in the form of $e^{-w_{min}}$; otherwise it will be impossible to satisfy $\alpha_t T_{t+1} < \delta/16$ for any arbitrarily small $\delta$ as $\|q(t)\| \to \infty$. The only function with such a property is the $\log(.)$ function. In fact, it turns out that $f$ must grow slightly slower than $\log(.)$ as we show next to satisfy (11), and to ensure the existence of a finite $t^*$ in Lemma 5.

Consider weight functions of the form $f(x) = \frac{\log(1+x)}{g(x)}$ where $g(x)$ is a strictly increasing function, chosen such that $f$ satisfies the conditions of Lemma 1. For example, by choosing functions that grow much slower than $\log(1+x)$, like $g(x) = \log(e + \log(1+x))$, we can make $f(x)$ behave approximately like $\log(1+x)$ for large ranges of $x$.

Assume $g(0) \geq 1$, then

$$f'(x) \leq \frac{1}{1+x}. \tag{15}$$

The inverse of $f$ cannot be expressed explicitly, however, it can be written as

$$f^{-1}(x) = \exp(xg(f^{-1}(x))) - 1. \tag{16}$$

Therefore,

$$\begin{aligned} f'(f^{-1}(w_{min}) - 1) &\leq \frac{1}{f^{-1}(w_{min})} && (17) \\ &= \frac{1}{\exp(w_{min} g(f^{-1}(w_{min}))) - 1}. && (18) \end{aligned}$$

Using (17), the conditions of Lemma 5 are satisfied if there exists a $q_{th}$ large enough such that

$$2N16^N \exp(4Nw_{max}) \frac{1}{\exp(w_{min} g(f^{-1}(w_{min}))) - 1} \leq \delta/16 \tag{19}$$

for $\|q(t)\| \geq q_{th}$.

Using (16) and noting that $w_{min} = \frac{\epsilon}{2N} w_{max}$, (19) can be written as

$$2N16^N \exp\left(w_{min}\left[\frac{8N^2}{\epsilon} - g(f^{-1}(w_{min}))\right]\right)\left(1 + \frac{1}{f^{-1}(w_{min})}\right) \leq \delta/16 \tag{20}$$

Consider fixed, but arbitrary, $N$ and $\epsilon$. As $q_{max} \to \infty$, $w_{max} \to \infty$, and consequently $w_{min} \to \infty$ and $f^{-1}(w_{min}) \to \infty$. Therefore, the exponent $\frac{8N^2}{\epsilon} - g(f^{-1}(w_{min}))$ is negative for $q_{max}$ large enough, and thus, there is a threshold $q_{th}$ such that for all $q_{max} > q_{th}$, the condition (20) is satisfied. To be more accurate, it suffices to choose

$$q_{th} = f^{-1}\left(\frac{2N}{\epsilon} \times \max\left\{\log(\frac{64N16^N}{\delta}), f(g^{-1}(\frac{16N^2}{\epsilon}))\right\}\right). \tag{21}$$

Then, it follows from Lemma 5 that $\|\pi_t - \mu_t\|_{TV} \leq \frac{\delta}{4}$, whenever $\|q(t)\| > q_{th} + t^*$.

**Remark 4.** *The assumption $g(0) \geq 1$ is not required, since, as we saw in the above analysis, only the asymptotic behavior of $g$ is important. If we choose $q_{th}$ large enough such that*

$$g(f^{-1}(w_{min}(t)) - 1) \geq 1 \tag{22}$$

*when $\|q(t)\| \geq q_{th}$, then (17) holds and the rest of the analysis follows exactly. In particular, in order to get an explicit formula for $f^{-1}$, we can choose $g(x) = \log(1+x)^{\theta}$ for some $0 < \theta < 1$. The weight function for such a $g$ is $f(x) = (\log(1+x))^{1-\theta}$, and $f^{-1}$ and has the closed form*

$$f^{-1}(x) = \exp(x^{\frac{1}{1-\theta}}) - 1.$$

*Then (21) yields*

$$q_{th} = \exp\left(\max\left\{\frac{2N}{\epsilon}\log(\frac{64N16^N}{\delta}), \frac{2N}{\epsilon}(\frac{16N^2}{\epsilon})^{\frac{1}{\theta}}\right\}^{\frac{1}{1-\theta}}\right). \tag{23}$$

*It is easy to check that for $q(t) \geq \exp\left((\frac{2N}{\epsilon})^{\frac{1}{1-\theta}}\log(1+e)\right)$, $w_{min}(t) \geq f(e)$ which satisfies (22). Therefore, obviously, (22) also holds for $q_{th}$ of (23).*

The last step of the proof is to determine the constant $B$ in (14), so we need to find $t^*$. Let $t_1$ be the first time that $q_{max}(t)$ hits $q_{th}$, then

$$\begin{aligned}
\sum_{k=t_1}^{t_1+t}\frac{1}{T_k^2} &\geq 16^{-2N}\sum_{k=t_1}^{t_1+t} e^{-8Nf(q_{max}(k))} \\
&= 16^{-2N}\sum_{k=t_1}^{t_1+t} e^{-8N\frac{\log(1+q_{max}(k))}{g(q_{max}(k))}} \\
&= 16^{-2N}\sum_{k=t_1}^{t_1+t}(1+q_{max}(k))^{-\frac{8N}{g(q_{max}(k))}} \\
&\geq 16^{-2N}\sum_{k=1}^{t}(1+q_{th}+k)^{-\frac{8N}{g(q_{th})}} \\
&\geq 16^{-2N}t(1+q_{th}+t)^{-\frac{8N}{g(q_{th})}}
\end{aligned}$$

Therefore, by Lemma 5, it suffices to find the smallest $t$ that satisfies

$$16^{-2N}t(1+q_{th}+t)^{-\frac{8N}{g(q_{th})}} \geq \log(4/\delta) + \frac{N}{2}\log(2(1+q_{th}))$$

for a threshold $q_{th}$ large enough satisfying (21). Recall that $g(.)$ is an increasing function, therefore, by choosing $q_{th}$ large enough, $\frac{8N}{g(q_{th})}$ can be made arbitrary small. Then a finite $t^*$

always exists since

$$\lim_{t^* \to \infty} t^*(1 + q_{th} + t^*)^{-\frac{8N}{g(q_{th})}} = \infty.$$

In particular, for the function $f(q) = (\log(1+q))^{1-\theta}$, $0 < \theta < 1$, and the choice of $q_{th}$ in (23), we have

$$\frac{8N}{g(q_{th})} = \frac{8N}{\log(1+q_{th})^{\theta}} < \frac{\epsilon}{2N}.$$

Note that

$$\frac{t}{(t+1+q_{th})^{\epsilon/2N}} \geq \frac{t^{1-\epsilon/2N}}{(2+q_{th})^{\epsilon/2N}}$$

and therefore, it is sufficient to choose $t^*$ to be

$$t^* = \left[(2+q_{th})^{\frac{\epsilon}{2N}} 16^N \log\left(\frac{4}{\delta}(2(1+q_{th}))^{N/2}\right)\right]^{\frac{1}{1-\frac{\epsilon}{2N}}} \tag{24}$$

This concludes the proof of the main Theorem.

### *E. Extension of the proof to the distributed implementation*

The distributed algorithm is based on multiple site-update (or parallel operating) Glauber dynamics as defined next. Consider the graph $G(V,E)$ as before and a weight vector $\tilde{W} = [\tilde{w}_1, \tilde{w}_2, ..., \tilde{w}_N]$. At each time $t$, a decision schedule $m(t) \subseteq \mathcal{M}$ is selected at random with positive probability $\alpha(m(t))$. Then, for all $i \in m(t)$,

(i) If $x_j(t-1) = 0$ for all nodes $j \in \mathcal{N}(i)$, then $x_i(t) = 1$ with probability $\frac{\exp(\tilde{w}_i)}{1+\exp(\tilde{w}_i)}$, or $x_i(t) = 0$ with probability $\frac{1}{1+\exp(\tilde{w}_i)}$ .
Otherwise, $x_i(t)$=0.

(ii) $x_j(t) = x_j(t-1)$ for all $j \notin m(t)$.

The Markov chain $X(t)$ is aperiodic and irreducible if $\cup_{m \in \mathcal{M}_0} = V$ (See [3] for more detail). Also, it can be shown that $X(t)$ is reversible, and it has the same stationary distribution as regular Glauber dynamics in (8). Here, we will assume that $\alpha_{min} := \min_m \alpha(m) \geq (1/2)^N$. Then, the mixing time of the chain is charachterized by the followng Lemma.

**Lemma 7.** *For the multiple site-update Glauber Dynamics with the weight vector $\tilde{W}$ on a graph $G(V,E)$ with $|V| = N$,*

$$T_t \leq \frac{64^N}{2} \exp(4N\tilde{w}_{max}). \tag{25}$$

*where $\tilde{w}_{max} = \max_{i \in V} \tilde{w}_i$.*

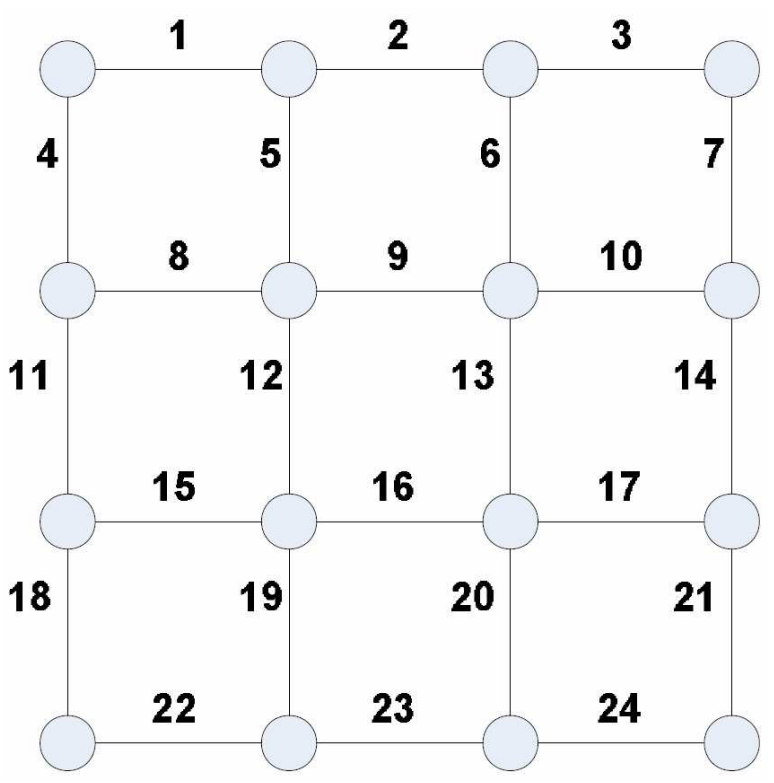

Fig. 1. A grid network with 24 links.

See the appendix for the proof. The distributed algorithm uses a time-varying version of the multiple-site update Glauber dynamics, where the weights change with time. Although the upperbound of Lemma 7 is loose, it is sufficient to prove the optimality of the algorithm. The analysis is the same as the argument for the basic algorithm. Let $D$ and $W$ denote the lengths of the data slot and the control slot. Thus, the distributed algorithm can achieve a fraction $\frac{D}{D+W}$ of the capacity region. In particular, recall the simple randomized machanism, in section III-B, where each node joins the decison schedule by sending an INTENT message with probability $1/2$. Note that in this case $\alpha_{min} \geq (1/2)^N$, and also it sufficies to allocate a short mini-slot at the begining of the slot for the purpose of control. By choosing the data slot to be much larger than the control slot, the algorithm can approach the full capacity.

## V. Simulation Results

In this section, we evaluate the performance of diffrent weight functions via simulations. For this purpose, we have considered the grid network of Figure 1, which has $16$ nodes and $24$ links, under one hop interference constraint. Consider the following maximal schedules

$$M_1 = \{1, 3, 8, 10, 15, 17, 22, 24\}$$

$$M_2 = \{4, 5, 6, 7, 18, 19, 20, 21\}$$

$$M_3 = \{1, 3, 9, 11, 14, 16, 22, 24\}$$

$$M_4 = \{2, 4, 7, 12, 13, 18, 21, 23\}$$

With a little abuse of abuse of notation, let $M_i$ also be a vector that its $i$-th element is $1$ if $i \in M_i$ and $0$ otherwise. We consider arrival rates that are a convex combination of the above maximal schedules scaled by $0 \leq \rho < 1$, e.g.,

$$\lambda = \rho \sum_{i=1}^{4} c_i M_i, \;\; c = [0.2, 0.3, 0.2, 0.3].$$

Note that, as $\rho \to 1$, $\lambda$ approaches a point on the boundary of the capacity region. We simulate the distributed algorithm, and use the following randomized mechanism, as in [3], similar to IEEE 802.11 DCF standard, to generate the decision schedules in the control slots. At time slot $t$:

1) Link $i$ selects a random back-off time $T_i$ uniformly in $[0, W-1]$ and waits for $T_i$ control mini-slots.
2) IF link $i$ hears an INTENT message from a link in $\mathcal{N}(i)$ before the $(T_i+1)$-th control mini-slot, $i$ will not be included in $m(t)$ and will not transmit an INTENT message anymore.
3) IF link $i$ does not hear an INTENT message from any link in $\mathcal{N}(i)$ before the $(T_i+1)$-th control mini-slot, it will broadcast an INTENT message at the beginning of the $(T_i+1)$-th control mini-slot. Then, if there is no collisions (i.e., no other link in $\mathcal{N}(i)$ transmits an INTENT message in the same mini-slot), link $i$ will be included in $m(t)$.

Once $m(t)$ is found, the access probabilities are determined as described in the distributed algorithm in section III-B. Here, we choose $W = 32$ (which is compatible with the back-off window size specified in IEEE 802.11 DCF).

In our simulations, the performance of $\log(1+x)$ and $\frac{\log(1+x)}{\log(e+\log(1+x))}$ is very close to each other, so in the plots, for brevity, we use the name $\log$ while the results actually belong to the function $\frac{\log(1+x)}{\log(e+\log(1+x))}$.

Figure 2 shows the average queue-length evolution (total queue-length divided by the number of links), for the weight functions $f(x) = \frac{\log(1+x)}{\log(e+\log(1+x))}$ and $f(x) = \log\log(e+x)$ and for loadings $\rho = 0.8$ and $0.82$. While both functions keep the queues stable, however as it is expected, the average-queue lengths for the weight function $\frac{\log}{\log\log}$ are much smaller than those for $\log\log$. Moreover, $\frac{\log}{\log\log}$ yields a faster convergence to the steady state. The performance gap of two functions, in terms of the average queue-length and the convergence speed, increases significantly for larger loadings; for example see Figure 3 for $\rho = 0.85$. Figures 4 and 5 show the delay performance (time-average queue-length per link) of the two weight functions under different loadings. As it is evident from the figures, $\log$ has a significantly smaller delay than what is incurred by using the weight $\log\log$. A natural question is that whether there exists a function faster than $\log$-wise functions that still stabilizes any general network. If such a function exists, then one will expect to get a better delay performance. Our conjecture is that, since the mixing time is, in general, exponential in $w_{max}$, $\log$ is the fastest weight function that can make the network change in an adiabatic manner, and hence keep the system close to its equilibrium (stationary distribution). We tried faster weight functions, such as $q$ and $\sqrt{q}$, but they resulted in unstable systems (for example see Figure 6).

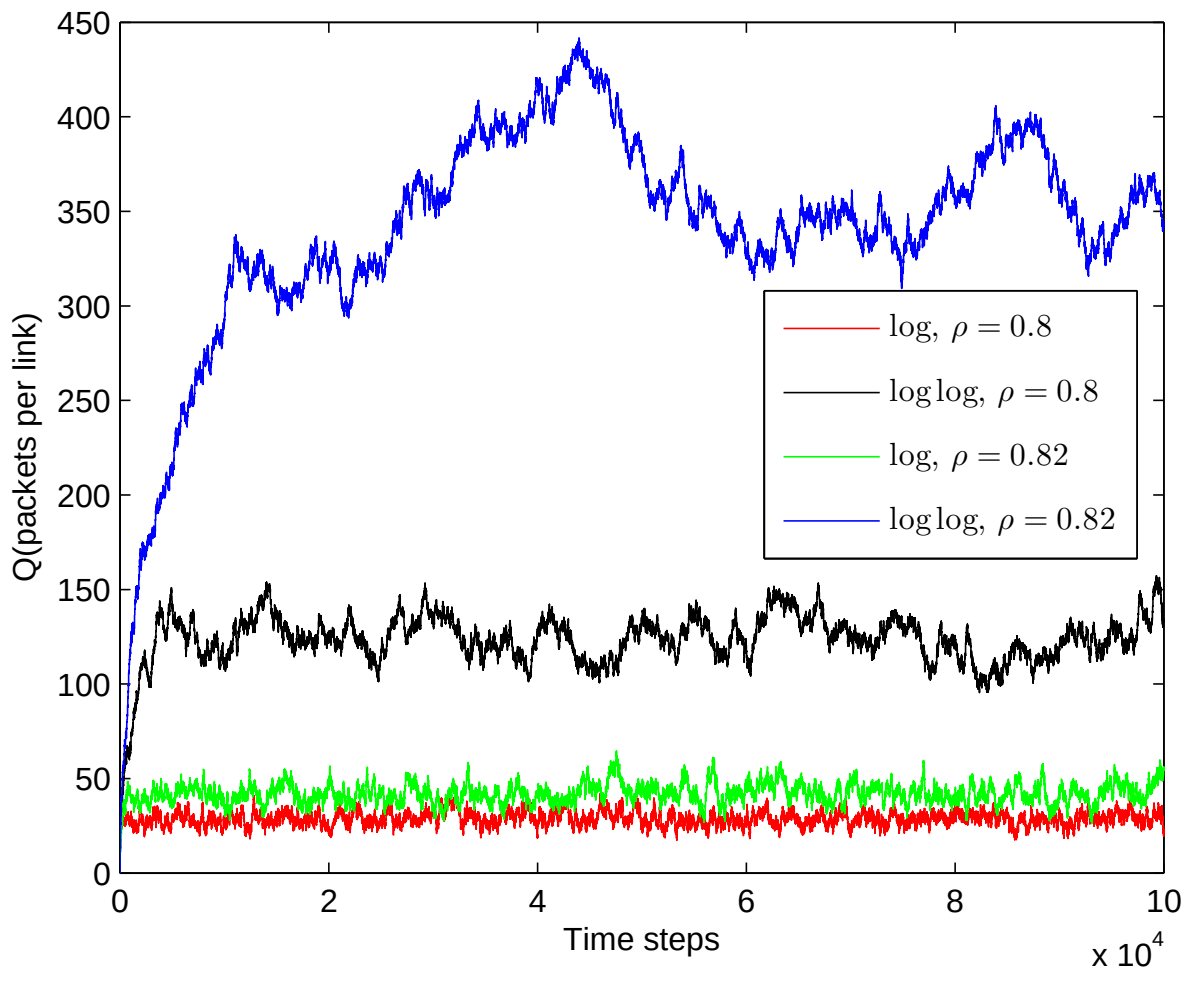

Fig. 2. The evolution of average queue-length for $\log\log$ and $\frac{\log}{\log\log}$(called $\log$ in the plots).

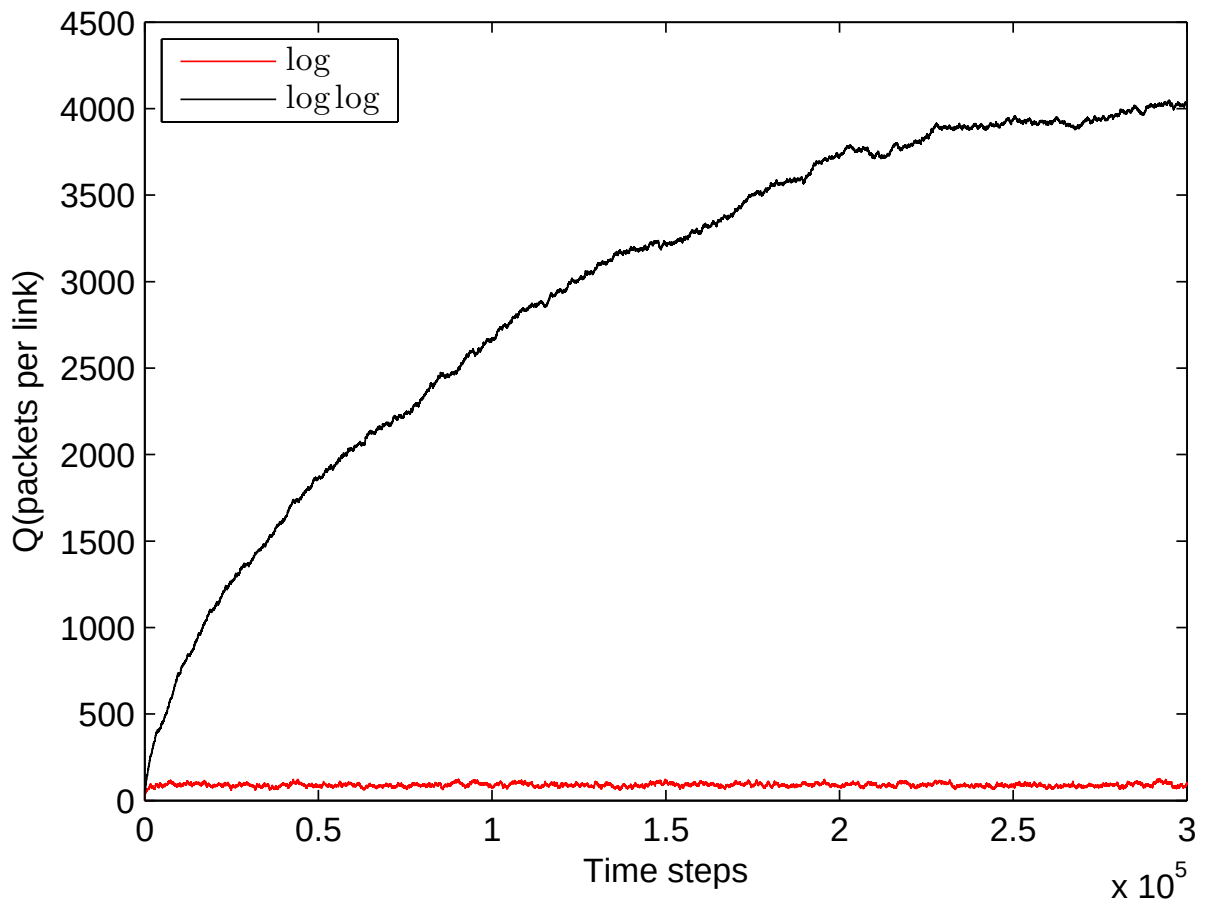

Fig. 3. The evolution of average queue-lengths for $\rho = 0.85$.

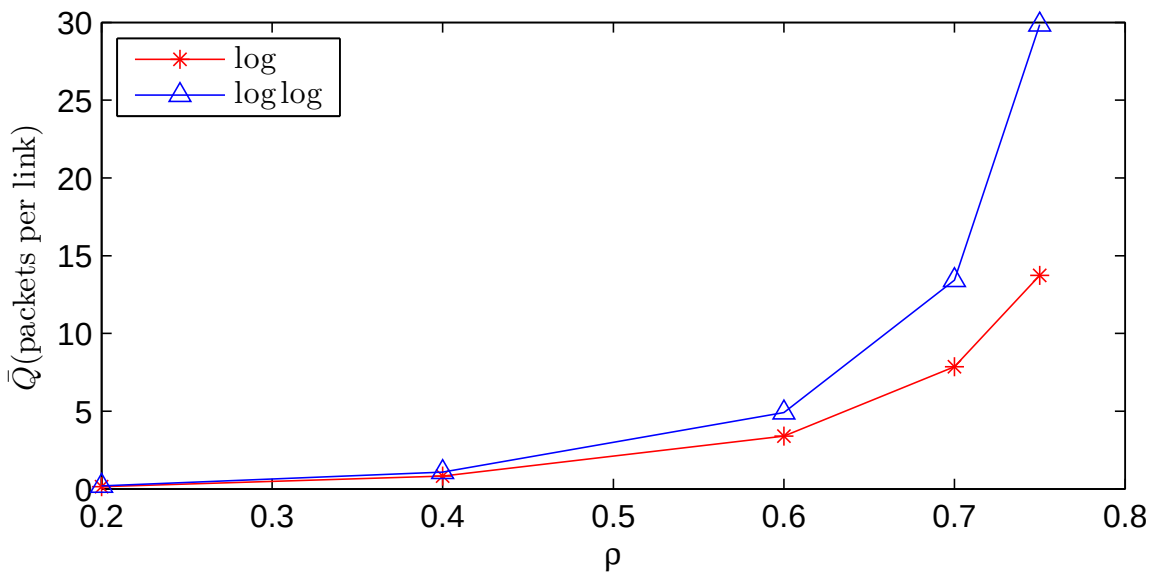

Fig. 4. Time-average of queue-length per link for low and moderate values of $\rho$.

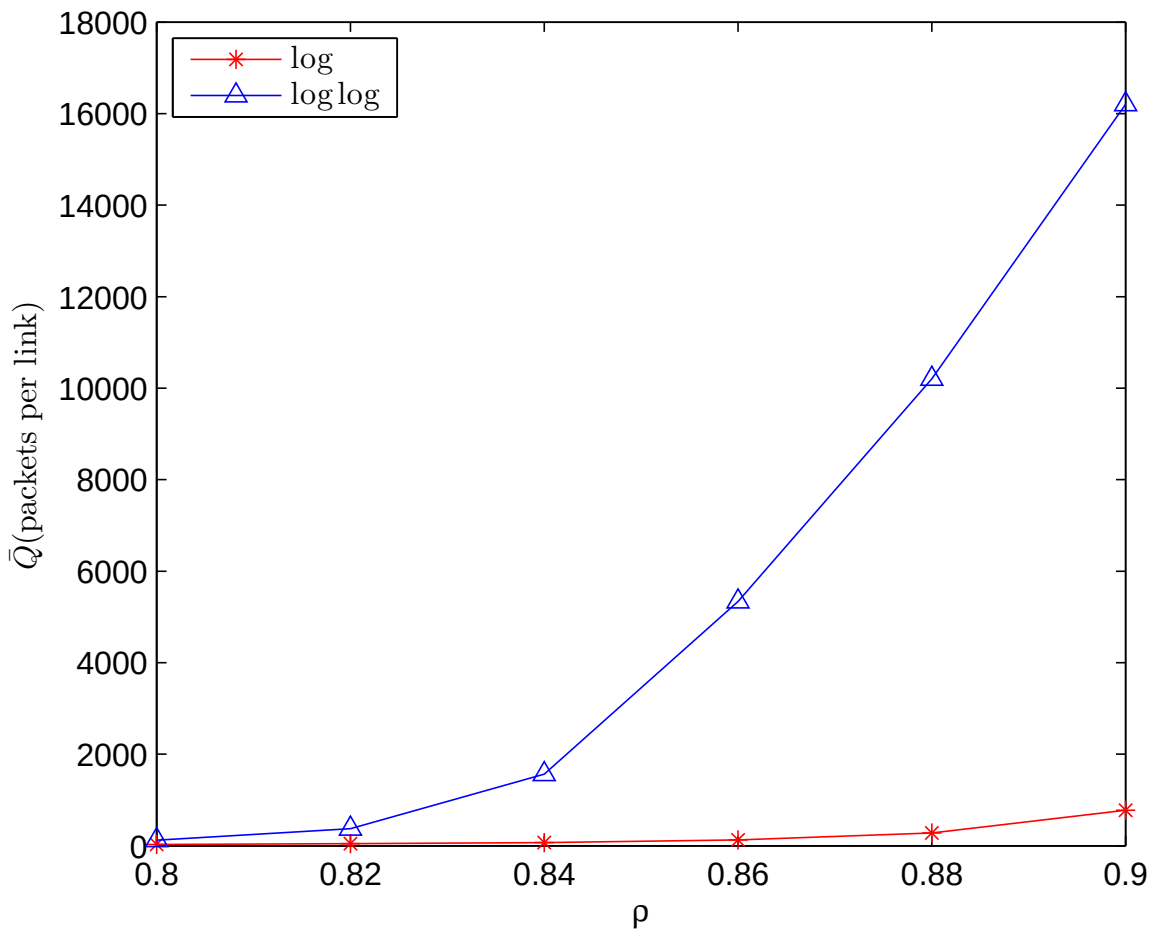

Fig. 5. Time-average of queue-length per link for high values of $\rho$.

## VI. Conclusions

In this paper, we considered the design of efficient CSMA algorithms that are throughput optimal and have a good delay performance. The algorithm is essentially a Glauber Dynamics with, potentially, multiple-site updates at each time-slot. Access probabilities depend on links weights, where the weight of each link is chosen to be an appropriate function of its queue-length. In particular, we showed that weight functions of the form $f(q) = \log(q)/g(q)$ yield throughput-optimality and low delay performance. The function $g(q)$ can grow arbitrarily slowly such that $f(q) \approx \log(q)$.

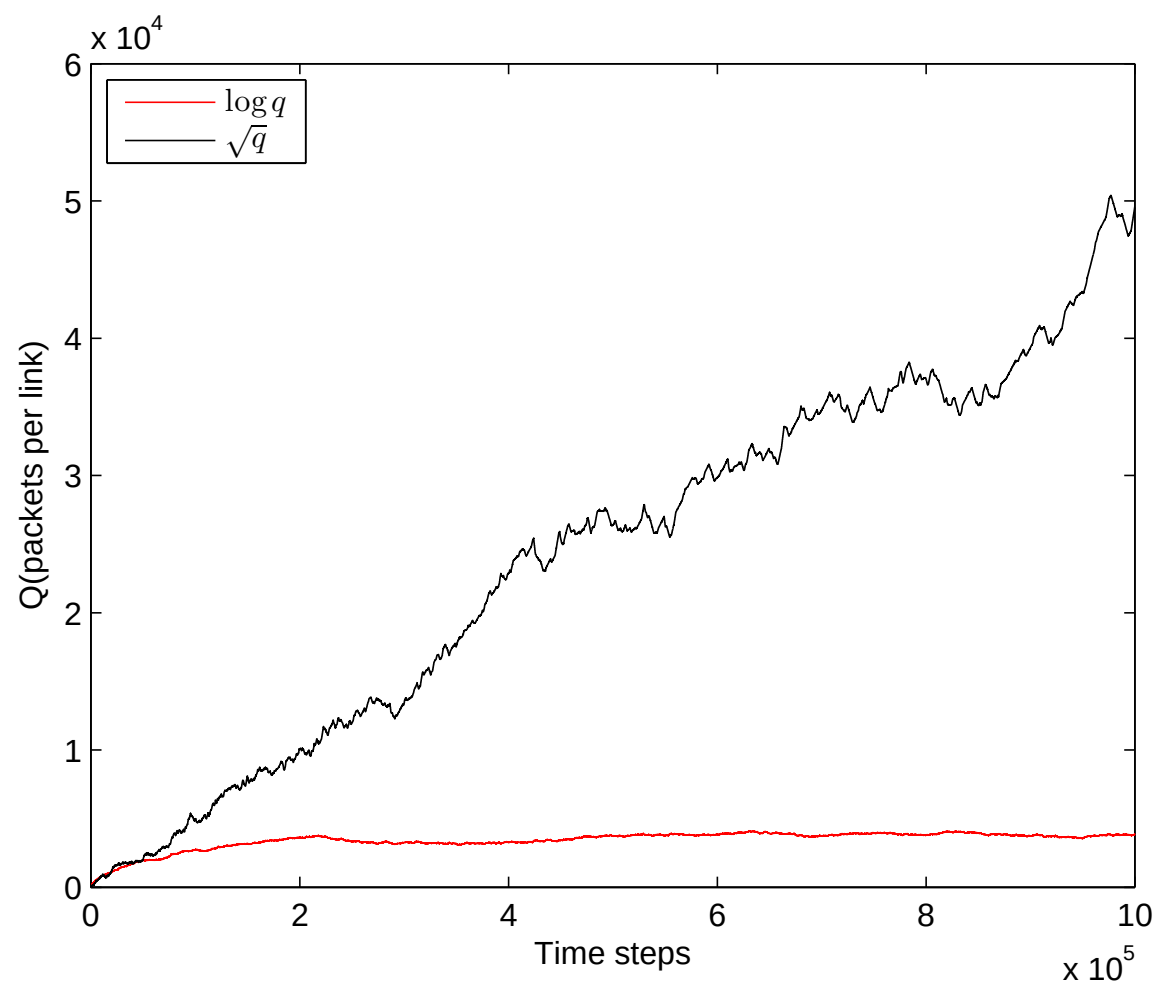

Fig. 6. The weight function $\sqrt{q}$ makes the system unstable ($\rho = 0.92$).

## ACKNOWLEDGMENT

The authors would like to thank Bo (Rambo) Tan, at the Coordinated Science Lab., for his assistance with simulations.

## APPENDIX A

*Proof of Lemma 1:* Let $f$ be strictly increasing with $f(0) = 0$, therefore $\lim_{q\to\infty} f(q) = \infty$. Consider an arbitrary $\epsilon > 0$ and $M_2 > 0$. We need to to find a $Q > 0$ such that for all $q > Q$,

$$f(q + M_2) \leq (1 + \epsilon)f(q)$$

or equivalently,

$$\frac{f(q + M_2) - f(q)}{f(q)} < \epsilon.$$

Since $f$ is concave[3],

$$\frac{f(q + M_2) - f(q)}{f(q)} \leq \frac{f'(q)}{f(q)} M_2.$$

It also follows from concavity that $f'$ is nonincreasing, and $f'(q) > 0$ for all $q > 0$ since $f$ is strictly increasing. Therefore $f'(q)$ must have a limit as $q \to \infty$ while $f(q)$ grows to infinity. Therefore, by choosing $q$ large enough, $\frac{f'(q)}{f(q)}$ can be made smaller than $\epsilon/M_2$ . This concludes the proof of the upper bound. The proof of the lower bound follows similarly. ■

[3]We have implicitly considered the class of continuously differentiable functions.

## Appendix B

*Proof of Lemma 3:* The upper-bound in Lemma 3 is based on the conductance bound [5], [6]. First, for a nonempty set $B \subset E$, define the followings:

$$\pi(B) = \sum_{i \in B} \pi(i),$$

$$F(B) = \sum_{i \in B, j \in B^c} \pi(i) p_{ij},$$

Then, conductance is defined as

$$\phi(P) = \inf_{B:\pi_t(B) \leq 1/2} \frac{F(B)}{\pi(B)}.$$

**Lemma 8.** *(Conductance Bounds)*

$$1 - 2\phi(P) \leq \lambda_2 \leq 1 - \frac{\phi^2(P)}{2}$$

The conductance can be further lower bounded as follows.

$$\begin{aligned} \phi(P) &= \inf_{B:\pi(B) \leq 1/2} \frac{\sum_{X \in B, Y \in B^c} \pi(X) P(X,Y)}{\pi(B)} \\ &\geq 2 \inf_{B \subseteq \mathcal{M}} \sum_{X \in B, Y \in B^c} \pi(X) P(X,Y) \\ &\geq 2 \min_{x} \pi(X) \min_{X \neq Y} P(X,Y) \end{aligned}$$

For our Glauber Dynamics, the stationary distribution is lower bounded by

$$\pi(\rho) \geq \frac{1}{\sum_{\rho} \exp(\sum_{i \in \rho} \tilde{w}_i)} \geq \frac{1}{|\mathcal{M}| \exp(N \tilde{w}_{max})}.$$

In addition, $X$ and $Y$ can differ in at only one site, and it is easy to see that

$$P(X,Y) \geq \frac{1}{N} \frac{1}{1 + \exp(\tilde{w}_{max})},$$

So

$$\begin{aligned} \phi(P) &\geq \frac{1}{N 2^{N-1} (1 + \exp(w_{max})) \exp(N w_{max})} \\ &\geq \frac{1}{N 2^N \exp((N+1) w_{max})}. \end{aligned}$$

Therefore,

$$\begin{aligned} \lambda_2(P) &\leq 1 - \frac{1}{2N^2 4^N \exp(2(N+1) w_{max})} \\ &\leq 1 - \frac{1}{16^N \exp(4N w_{max})}. \end{aligned}$$

By Gershgorin's theorem (e.g. see the appendix of [6]), for a stochastic matrix $[P_{ij}]$,

$$\lambda \geq -1 + 2\min P_{ii}.$$

For our Glauber-Dynamics,

$$\begin{aligned} P_{\rho\rho} &\geq 1 - \frac{1}{N}\sum_{i\in\rho}\frac{1}{1+\exp(\widetilde{w}_i)} - \frac{1}{N}\sum_{i\in V\setminus\rho}\frac{\exp(\widetilde{w}_i)}{1+\exp(\widetilde{w}_i)} \\ &\geq 1 - \frac{1}{N}\sum_{i=1}^{N}\frac{\exp(w_{max})}{1+\exp(w_{max})} \\ &= \frac{1}{1+\exp(w_{max})}. \end{aligned}$$

So,

$$\lambda_r \geq -1 + \frac{2}{1+\exp(w_{max})} = \frac{1-\exp(w_{max})}{1+\exp(w_{max})}.$$

Therefore,

$$\max\{\lambda_2, |\lambda_r|\} = \lambda_2$$

and the SLEM of $P$ is upperbounded by

$$\sigma \leq 1 - \frac{1}{16^N \exp(4Nw_{max})}. \tag{26}$$

Consequently

$$T \leq 16^N \exp(4Nw_{max}). \tag{27}$$

■

## APPENDIX C

*Proof of Lemma 4:*

The corresponding stationary distributions at times $t$ and $t+1$ are respectively given by

$$\pi_t(\rho) = \frac{1}{Z}\exp(\sum_{i\in\rho}\widetilde{w}_i(t)),$$

and,

$$\pi_{t+1}(\rho) = \frac{1}{Z}\exp(\sum_{i\in\rho}\widetilde{w}_i(t+1)),$$

So

$$\frac{\pi_{t+1}(\rho)}{\pi_t(\rho)} = \exp(\sum_{i\in\rho}\widetilde{w}_i(t+1) - \widetilde{w}_i(t))$$

or

$$\exp\left(-\sum_{i\in\rho}|\widetilde{w}_i(t+1)-\widetilde{w}_i(t)|\right)\leq\frac{\pi_{t+1}(\rho)}{\pi_t(\rho)}\leq\exp\left(\sum_{i\in\rho}|\widetilde{w}_i(t+1)-\widetilde{w}_i(t)|\right).$$

Let $q_t^*$ denote $f^{-1}(w_{min}(t))$, and $\widetilde{q}(t)=\max\{q_t^*,q(t)\}$, where $q(t)$ is the vector of queue lengths at time $t$. Recall that $f$ is a concave and increasing function. Hence, if $\widetilde{q}_i(t+1)\geq\widetilde{q}_i(t)$,

$$\widetilde{w}_i(t+1)-\widetilde{w}_i(t)=f(\widetilde{q}_i(t+1))-f(\widetilde{q}_i(t))\leq f'(\widetilde{q}_i(t))(\widetilde{q}_i(t+1)-\widetilde{q}_i(t))\leq f'(\widetilde{q}_i(t)).$$

(note that $q_i(t+1)$ and $q_i(t)$ at most differ by one since there can at most one packet arrival or departure in a time slot). Similarly if $\widetilde{q}_i(t+1)\leq\widetilde{q}_i(t)$, $f(\widetilde{q}_i(t))-f(\widetilde{q}_i(t+1))\leq f'(\widetilde{q}_i(t+1))$. So

$$|\widetilde{w}_i(t+1)-\widetilde{w}_i(t)|\leq f'(\widetilde{q}_i(t))+f'(\widetilde{q}_i(t+1))\leq f'(\widetilde{q}_i(t+1)-1)+f'(\widetilde{q}_i(t+1))$$

and therefore

$$|\widetilde{w}_i(t+1)-\widetilde{w}_i(t)|\leq 2f'(q^*(t+1)-1).$$

Define

$$\alpha_t=2Nf'(q^*(t+1)-1), \tag{28}$$

then

$$e^{-\alpha_t}\leq\frac{\pi_{t+1}(\rho)}{\pi_t(\rho)}\leq e^{\alpha_t}. \tag{29}$$

The drift in $\pi_t$ is given by

$$\begin{aligned}\|\pi_{t+1}-\pi_t\|^2_{1/\pi_{t+1}} &= \|\frac{\pi_t}{\pi_{t+1}}-1\|^2_{\pi_{t+1}}\\ &= \sum_\rho \pi_{t+1}(\rho)(\frac{\pi_t(\rho)}{\pi_{t+1}(\rho)}-1)^2\\ &\leq \sum_\rho \pi_{t+1}(\rho)\max\{(e^{\alpha_t}-1)^2,(1-e^{-\alpha_t})^2\}\\ &\leq \max\{(e^{\alpha_t}-1)^2,(1-e^{-\alpha_t})^2\}\\ &= (e^{\alpha_t}-1)^2\end{aligned}$$

for $\alpha_t<1$. Thus,

$$\|\pi_{t+1}-\pi_t\|_{1/\pi_{t+1}}\leq 2\alpha_t \tag{30}$$

for $\alpha_t<1$, where

$$\alpha_t=2Nf'(f^{-1}(w_{min}(t+1))-1). \tag{31}$$

The distance between the true distribution and the stationary distribution at time $t$ can be bounded as follows. First, by triangle inequality,

$$\begin{aligned}\|\mu_t - \pi_t\|_{1/\pi_t} &\leq \|\mu_t - \pi_{t-1}\|_{1/\pi_t} + \|\pi_{t-1} - \pi_t\|_{1/\pi_t} \\ &\leq \|\mu_t - \pi_{t-1}\|_{1/\pi_t} + 2\alpha_{t-1}.\end{aligned}$$

On the other hand,

$$\begin{aligned}\|\mu_t - \pi_{t-1}\|^2_{1/\pi_t} &= \sum_\rho \frac{1}{\pi_t(\rho)}(\mu_t(\rho) - \pi_{t-1}(\rho))^2 \\ &= \sum_\rho \frac{\pi_{t-1}(\rho)}{\pi_t(\rho)} \frac{1}{\pi_{t-1}(\rho)}(\mu_t(\rho) - \pi_{t-1}(\rho))^2 \\ &\leq e^{\alpha_{t-1}} \|\mu_t - \pi_{t-1}\|^2_{1/\pi_{t-1}}.\end{aligned}$$

Therefore, for $\alpha_t < 1$,

$$\|\frac{\mu_t}{\pi_t} - 1\|_{\pi_t} \leq (1 + \alpha_{t-1})\|\mu_t - \pi_{t-1}\|_{1/\pi_{t-1}} + 2\alpha_{t-1}.$$

Suppose $\alpha_t \leq \delta/16$, then $\|\frac{\mu_t}{\pi_t} - 1\|_{\pi_t} \leq \delta/2$ holds for $t > t^*$, if

$$\|\mu_t - \pi_{t-1}\|_{1/\pi_{t-1}} \leq \delta/4$$

for all $t > t^*$.

Define $a_t = \|\mu_{t+1} - \pi_t\|_{1/\pi_t}$. Then

$$\begin{aligned}a_{t+1} &= \|\mu_{t+2} - \pi_{t+1}\|_{1/\pi_{t+1}} \\ &= \|\mu_{t+1}P_{t+1} - \pi_{t+1}\|_{1/\pi_{t+1}} \\ &\leq \sigma_{t+1}\|\mu_{t+1} - \pi_{t+1}\|_{1/\pi_{t+1}}\end{aligned}$$

where $\sigma_{t+1}$ is the SLEM of $P_{t+1}$, since $(P_{t+1}, \pi_{t+1})$ is reversible. Therefore,

$$a_{t+1} \leq \sigma_{t+1}[(1 + \alpha_t)a_t + 2\alpha_t]$$

Suppose $a_t \leq \delta/4$. Defining $T_t = \frac{1}{1-\sigma_t}$, we have

$$a_{t+1} \leq (1 - \frac{1}{T_{t+1}})[\delta/4 + (2 + \delta/4)\alpha_t].$$

Thus, $a_{t+1} \leq \delta/4$, if

$$(2 + \delta/4)\alpha_t < \frac{1}{T_{t+1}}(\delta/4 + (2 + \delta/4)\alpha_t),$$

or equivalently if

$$\alpha_t < \frac{\frac{\delta/4}{T_{t+1}}}{(2+\delta/4)(1-1/T_{t+1})}.$$

But

$$\frac{\frac{\delta/4}{T_{t+1}}}{(2+\delta/4)(1-1/T_{t+1})} > \frac{\frac{\delta/4}{T_{t+1}}}{4(1-1/T_{t+1})} > \frac{\delta}{16}\frac{1}{T_{t+1}},$$

so, it is sufficient to have

$$\alpha_t T_{t+1} \leq \delta/16.$$

Therefore, if there exists a time $t^*$ such that $a_{t^*} \leq \delta/4$, then $a_t \leq \delta/4$ for all $t \geq t^*$. To find $t^*$, note that $a_t > \delta/4$ for all $t < t^*$. So, for $t < t^*$, we have

$$\begin{aligned}
a_t &\leq (1-\frac{1}{T_t})[(1+\alpha_{t-1})a_{t-1} + 2\alpha_{t-1}] \\
&\leq (1-\frac{1}{T_t})[(1+\alpha_{t-1})a_{t-1} + 2\alpha_{t-1}4\frac{a_{t-1}}{\delta}] \\
&\leq (1-\frac{1}{T_t})(1+\alpha_{t-1}+\frac{8}{\delta}\alpha_{t-1})a_{t-1} \\
&\leq (1-\frac{1}{T_t})(1+\frac{\delta/16}{T_t}(1+\frac{8}{\delta}))a_{t-1} \\
&\leq (1-\frac{1}{T_t})(1+\frac{1}{T_t})a_{t-1} \\
&= (1-\frac{1}{T_t^2})a_{t-1} \\
&\leq e^{-\frac{1}{T_t^2}}a_{t-1}
\end{aligned}$$

Thus,

$$a_t \leq a_0 e^{-\sum_{k=1}^{t^*}\frac{1}{T_k^2}},$$

where

$$\begin{aligned}
a_0 &= \|\frac{\mu_1}{\pi_0} - 1\|_{\pi_0} \\
&= \|\mu_0 P_0 - \pi_0\|_{1/\pi_0} \\
&\leq \sigma(P_0)\|\mu_0 - \pi_0\|_{1/\pi_0} \\
&\leq \sqrt{\frac{1}{\pi_0^{min}}}
\end{aligned}$$

and

$$\begin{aligned}\pi_0^{min} &= \min_\rho \pi_0(\rho) \\ &\geq \frac{1}{\sum_\rho \exp(\sum_{i\in\rho}\widetilde{w}_i(0))} \\ &\geq \frac{1}{|\mathcal{M}|\exp(Nw_{max}(0))}\end{aligned}$$

which yields

$$a_0 \leq (2e^{w_{max}(0)})^{N/2}.$$

Putting everything together, $t^*$ must satisfy

$$(2e^{w_{max}(0)})^{N/2} e^{-\sum_{k=1}^{t^*}\frac{1}{T_k^2}} \leq \delta/4$$

or as a sufficient condition,

$$\sum_{k=1}^{t^*}\frac{1}{T_k^2} \geq \log(4/\delta) + N(w_{max}(0) + \log 2)/2.$$

■

## APPENDIX D

*Proof of Lemma 7:*

Let $\mathcal{M}_0 \subseteq \mathcal{M}$ be the set of all possible decision schedules. Given $X(t) = X$, for some $X \in \mathcal{M}$, the next state/schedule could be $X(t+1) = Y$ with the following transition probability

$$P(X,Y) = \sum_{m\in\mathcal{M}_0 : X\Delta Y\subseteq m} \alpha(m) \prod_{i\in m\setminus(Y\cup\mathcal{N}(X\cup Y))} \frac{1}{1+\exp(\tilde{w}_i)} \prod_{j\in m\cap Y} \frac{\exp(\tilde{w}_j)}{1+\exp(\tilde{w}_j)}, \quad (32)$$

where $X\Delta Y = (X\setminus Y)\cup(Y\setminus X)$.

The upper-bound in Lemma 7 is based on the conductance bound as in the proof of Lemma 3. Recall that the conductance can be lower bounded as follows.

$$\phi(P) \geq \; 2\min_X \pi(X)\min_{X\neq Y} P(X,Y)$$

As in the regular Glauber Dynamics,

$$\pi(X) \geq \frac{1}{2^N \exp(Nw_{max})}.$$

and,

$$P(X,Y) \geq \alpha_{min} \left( \frac{1}{1+\exp(w_{max})} \right)^N,$$

where $\alpha_{min} = \min_{m \in M_0} \alpha(m) \geq \frac{1}{2^N}$. Hence,

$$\begin{aligned} \phi(P) &\geq \frac{2}{4^N (1+\exp(w_{max}))^N \exp(N w_{max})} \\ &\geq \frac{2}{8^N \exp(2N w_{max})}. \end{aligned}$$

Therefore, based on the conductance upperbound,

$$\lambda_2(P) \leq 1 - \frac{2}{64^N \exp(4N w_{max})}$$

and by Gershgorin's theorem,

$$\lambda_r \geq -1 + \frac{2}{2^N (1+\exp(w_{max}))^N}.$$

Therefore,

$$\max\{\lambda_2, |\lambda_r|\} = \lambda_2$$

and the SLEM of $P$ is upperbounded by

$$\sigma_t \leq 1 - \frac{2}{64^N \exp(4N w_{max})},$$

Consequently

$$T \leq \frac{64^N}{2} \exp(4N w_{max}). \tag{33}$$

■